\documentclass[aps,nofootinbib,twocolumn,showpacs]{revtex4-1}
\usepackage{amsmath}
\usepackage{amssymb}
\usepackage{graphicx}

\begin{document}

\title{Comment on ``Nucleon spin-averaged forward virtual 
Compton tensor at large $Q^2$''}
\author{Michael C. Birse and Judith A. McGovern}
\affiliation{Theoretical Physics Division, School of Physics and Astronomy,\\
The University of Manchester, Manchester, M13 9PL, UK\\}

\begin{abstract}
In recent work, Hill and Paz apply the operator product expansion 
to forward doubly virtual Compton scattering. The resulting 
large-$Q^2$ form of the amplitude $W_1(0,Q^2)$ is compatible with 
the one we obtain by extrapolation of low-$Q^2$ results
from a chiral effective field theory, providing support for our
approach. That paper also presents a result for the two-photon
contribution to the Lamb shift in muonic hydrogen that has a much 
larger uncertainty than in previous work. We show that this an
overestimate arising from the inclusion of the proton pole term in 
the subtracted dispersion relation for $W_1$.

\end{abstract}

\maketitle

\vskip 10pt

In Ref.~\cite{hp17}, Hill and Paz use the operator product 
expansion (OPE) to determine the large-$Q^2$ behaviour of the amplitude 
for forward doubly virtual Compton scattering (VVCS) from a nucleon. 
This work extends and corrects an older treatment by Collins 
\cite{col79}. The result is important because it can help 
to constrain the uncertainty in the proton radius extracted from 
measurements of the Lamb shift of muonic hydrogen \cite{pohl10,ant13s}. 

In particular, the result provides the large-$Q^2$ form of the 
subtraction term in one of the dispersion relations used in 
that analysis \cite{pach99}. (See Ref.~\cite{ant13} for a summary 
of the theory of the Lamb shift in muonic hydrogen.) Previous work 
by the current authors \cite{bmcg12} used chiral effective field theory 
($\chi$EFT) to determine the low-$Q^2$ form of the subtraction term
and then extrapolated this to higher $Q^2$ by matching it on to
the asymptotic $1/Q^2$ behaviour of the OPE. 
The resulting values of the coefficient of the $1/Q^2$ term 
are compatible with those from the QCD calculations of
Ref.~\cite{hp17}, providing support for the approach 
taken in Ref.~\cite{bmcg12}.

Ref.~\cite{hp17} also presents results for the two-photon
contribution to the muonic Lamb shift obtained with a simple interpolation between their OPE result at large $Q^2$ and 
the low-energy theorem (LET) governing the behaviour as
$Q^2\rightarrow 0$. This gives an energy that is compatible with
the previous result, but with an uncertainty that is an 
order of magnitude larger. As we discuss below, this is an 
overestimate, arising in the most part from the way that the
authors of  Ref.~\cite{hp17} treat the pole terms of 
the Born amplitude.

To recap, most of the two-photon contributions to the muonic 
Lamb shift can be determined from empirical input using the methods 
outlined by Pachucki \cite{pach99}. The Born (elastic) pieces can be written as integrals over proton electromagnetic form factors. 
Dispersion relations can be used to express the inelastic pieces 
in terms of integrals over structure functions. 
However the dispersion relation for $W_1(\nu,Q^2)$ 
(the amplitude multiplying $-g_{\mu\nu}+q_\mu q_\nu/q^2$ in the 
usual tensor decomposition) requires a subtraction. The resulting 
term in the energy has the form of an integral over $W_1(0,Q^2)$.
This zero-energy limit of the amplitude for forward VVCS 
cannot be measured directly and is one of the main sources of uncertainty in extractions of the proton radius. Its small $Q^2$
behaviour is constrained by an LET \cite{skk96,dkms97,fs98} 
and can be calculated in $\chi$EFTs \cite{bmcg12,np08,alp14,pp15}.

The large-$Q^2$ behaviour of $W_1$ can be obtained from QCD using 
the OPE \cite{col79}. The calculation in Ref.~\cite{hp17}
shows that $Q^2W_1(0,Q^2)/(2M_p^2)\sim 0.27$--$0.37$ for
$Q\gtrsim 5$~GeV$^2$.
For comparison, the extrapolation of $\chi$EFT in Ref.~\cite{bmcg12}
gives a central value of $1.3$, but with a wide uncertainty
band: $0.2$--23. Although the central value differs from the OPE 
by a factor of 3 to 4, these results are consistent within that 
large uncertainty. Moreover, in the corresponding contribution to
the Lamb shift, the integral over $W_1$ is heavily weighted to low 
$Q^2$ so that differences at high-$Q^2$ have only a small effect. 
This indicates that that a smooth interpolation between 
$\chi$EFT and the OPE should lead to results that lie within the uncertainties of Ref.~\cite{bmcg12}. For example, even reducing the entire contribution from $Q^2>0.3$~GeV$^2$ by a factor of 4 would 
alter the Lamb shift by only about 0.3~$\mu$eV. 
This is not to imply that the extrapolation is an alternative determination
of the high-$Q^2$ behaviour in QCD, merely that it provides a good enough 
estimate in the context of the muonic Lamb shift.

The bottom line of Ref.~\cite{hp17} for the two-photon contribution to
the muonic Lamb shift is $\Delta E(2P{-}2S)=+30\pm 13$~$\mu$eV. 
For comparison, the result of Ref.~\cite{bmcg12} is $+33\pm 2$~$\mu$eV.
The central values are similar\footnote{There are minor differences 
in the input data. Ref.~\cite{hp17} takes the magnetic 
polarisability $\beta_M$ from the PDG \cite{pdg16} and a dipole 
ansatz for the elastic form factors. Ref.~\cite{bmcg12} takes $\beta_M$ from a $\chi$EFT analysis \cite{gmpf12} and the elastic 
contributions from the work of Carlson and Vanderhaeghen \cite{cv11a}, 
who considered several empirical parametrisations of the form factors. 
Both works take the inelastic contribution from Ref.~\cite{cv11a}.} 
but the uncertainty in Ref.~\cite{hp17} is a factor of 6 larger.

The main reason for this large uncertainty is 
that the subtraction is applied to a dispersion relation that includes 
the proton poles \cite{hp11,hp17}, in contrast to other treatments 
that use a dispersion relation for the amplitude with 
the Born terms removed \cite{pach99,cv11a,bmcg12}. If the proton form 
factors were known with sufficient accuracy, the results from both approaches would be the same, but this is not the case in practice.
The version used in Ref.~\cite{hp17}
leads to integrals that are much more sensitive to proton form
factors that are not currently well determined.

At low $\nu$ and $Q^2$, the forward VVCS amplitude can be written 
in the form \cite{skk96,dkms97,fs98,bmcg12}\footnote{Note that 
$W_1$ as defined in Ref.~\cite{hp17} differs from $T_1$ in 
Ref.~\cite{bmcg12} by a factor of $2M/e^2$.} 
\begin{eqnarray}
W_1(\nu,Q^2)&=&\frac{2Q^4}{Q^4-4M^2\nu^2}\,G_M^2(Q^2)
-2F_D^2(Q^2)\cr
\noalign{\vspace{5pt}}
&&+\frac{2M}{\alpha}\left[Q^2\beta_M
+\nu^2(\alpha_E+\beta_M)\right]+\cdots,\quad
\label{eq:w1}
\end{eqnarray}
up to terms of fourth order in $\nu$ and $Q$. Here $\alpha_E$ and
$\beta_M$ are the electric and magnetic polarisabilities of the
proton \cite{gmpf12}, and 
\begin{equation}
F_D(Q^2)=\frac{G_E(Q^2)+\frac{Q^2}{4M^2}\,G_M(Q^2)}{1+\frac{Q^2}{4M^2}}
\end{equation} 
is its Dirac form factor. The non-pole term involving $F_D^2$ 
(the second term in Eq.~(\ref{eq:w1})) has been 
the subject of some controversy and we return to it below; 
for now we focus on the pole term.

Setting $\nu=0$ and expanding $W_1$ to order $Q^2$ gives the LET 
in the form in Refs.~\cite{hp11,hp17}:
\begin{eqnarray}
W_1(0,Q^2)&=&2\kappa(1+\kappa)
-\frac{2}{3}\,(1+\kappa)^2r_M^2\,Q^2\cr
\noalign{\vspace{5pt}}
&&+\frac{2}{3}\,r_E^2\,Q^2-\frac{\kappa}{M^2}\,Q^2
+\frac{2M}{\alpha}\,\beta_M\,Q^2+\cdots,\quad
\label{eq:w1let}
\end{eqnarray}
where $r_E$ and $r_M$ are the charge and magnetic radii of the proton and $\kappa$ is its anomalous magnetic moment. The slope of this with respect to $Q^2$ controls the low-$Q^2$ contribution to the 
subtraction term (cf.\ Eq.~(37) of Ref.~\cite{hp17}). 
Using the same input as in Ref.~\cite{hp17}, the various pieces 
of the LET contribute to the slope
as follows: Born pole (second term of Eq.~(\ref{eq:w1let})) 
$-80.4\pm 7.9$~GeV$^{-2}$, 
Born non-pole (third and fourth terms) $+11.1\pm 0.2$~GeV$^{-2}$, 
magnetic polarisability, $+8.4\pm 1.3$~GeV$^{-2}$. 

As noted above, the dispersion relation used in Ref.~\cite{hp17}
includes the Born pole. All three pieces listed above contribute to 
its subtraction. In this case, both the slope of $W_1$ and 
its error are dominated by the subtraction of the pole term.
The rather poorly determined magnetic radius of the proton 
appears multiplied by a large factor containing the square of 
the magnetic moment. It thus contributes significantly to  
the large uncertainty found in Ref.~\cite{hp17}.

In addition, the subtraction of the Born pole leads to a second 
enhancement in the uncertainty. The slope of the corresponding 
term in Eq.~(\ref{eq:w1let}) is an order of magnitude larger than 
that arising from the magnetic polarisability. Including it leads to 
a large factor multiplying the poorly-known form factor 
of the subtraction term. Even with the constraints from $\chi$EFT,
this form factor is one of the main contributions to the uncertainty 
in Ref.~\cite{bmcg12}. In their interpolation between the 
low-$Q^2$ regime and their OPE, the authors of Ref.~\cite{hp17} 
use no theoretical input on terms of higher-order in $Q^2$, 
assuming that the order-$Q^4$ term has a typical hadronic scale 
but leaving even its sign unfixed. This leads to a somewhat larger 
relative error on their result for the subtraction term than 
that in Ref.~\cite{bmcg12}. Multiplying this by the slope of $W_1$ 
including the pole contribution
enhances the absolute error by nearly an order of magnitude. 
Combined with the contribution from the magnetic radius
just discussed, this leads to the large overall uncertainty in the 
two-photon energy found in Ref.~\cite{hp17}. In contrast, 
if the Born pole is removed from the dispersion relation, 
the magnetic radius does not appear separately from the magnetic form factor
in the Born contribution to the Lamb shift. This form factor has a much 
better determined dependence on $Q^2$ and so leads to a smaller 
uncertainty.

The Born pole is a well defined nonanalytic structure in the VVCS amplitude, 
with a residue that is given in terms of on-shell proton form factors. 
It is thus straightforward to remove it from the 
amplitude $W_1$ and apply a subtracted dispersion relation to the 
remainder \cite{pach99,cv11a}. The large uncertainties in 
Ref.~\cite{hp17} associated with the pole contribution to 
$W_1(0,Q^2)$ are absent in approaches that use a dispersion 
relation with the pole removed \cite{pach99,cv11a,bmcg12}.
In addition, the low-$Q^2$ form factor for the subtraction
has been calculated using $\chi$EFT in Ref.~\cite{bmcg12}, including 
the order-$Q^4$ term that is left undetermined in Ref.~\cite{hp17}. 
This further constrains the subtraction term in the region that 
makes an important contribution to Lamb shift.

As just discussed, the main uncertainty in their two-photon exchange 
energy is driven by the way the pole term in the VVCS amplitude
is treated in Ref.~\cite{hp17}. There is one quite separate 
final issue, which concerns the treatment of the non-pole Born 
term (the $F_D^2$ term in Eq.~(\ref{eq:w1})). This can be 
generated from a Dirac equation with proton form factors --
the procedure that the authors of Ref.~\cite{hp17} refer to as the 
``sticking-in form factors" ansatz. However, 
as argued in Ref.~\cite{bmcg12}, the non-pole Born term follows
from Lorentz invariance and so it is natural to treat it together
with the pole term. Up to order $Q^2$, the terms in $F_D(Q^2)^2$
are required by low-energy theorems, but those of order $Q^4$ and
above can get contributions from low-energy constants that are beyond 
the order of the $\chi$EFT calculation in Ref.~\cite{bmcg12}. 
Using the empirical form factor for $F_D(Q^2)^2$  instead of the one 
from the $\chi$EFT means that some higher-order terms have been 
included in $W_1(0,Q^2)$. 
Provided they are of natural size, such terms have already been accounted 
for in the estimated uncertainty on the $\chi$EFT result \cite{bmcg12} and 
the effects of this choice should fall within that error estimate. 

In Appendix B of ref.~\cite{hp17}, it is suggested that there may
be larger uncertainties associated with this treatment. 
This is illustrated by a comparison of
results for the subtraction term from the third order $\chi$PT result of 
Ref.~\cite{alp14} and the fourth-order EFT-informed result of \cite{bmcg12}. 
This is misleading since, as Alarc\'on \textit{et al.}~point out, 
their subtraction term cannot not be compared with those from other 
approaches. This is because they find a cancellation between 
contributions of the $\Delta$ to the subtraction and inelastic terms, 
and hence they omit these entirely from both terms \cite{alp14}.
In particular, their result---if taken as complete---would correspond to
a magnetic polarisability with the wrong sign.
At fourth order the $\chi$EFT has the freedom to reproduce the experimental 
value of $\beta$, and so the subtraction term has the correct slope in the small 
$Q^2$ region that is crucial for the Lamb shift~\cite{bmcg12}.

\section*{Acknowledgments}

We are grateful to R. Hill and G. Paz for helpful discussions 
clarifying their approach. This work was supported by the UK 
STFC under grant ST/L005794/1.

\end{document}